\documentclass[12pt]{iopart}
\usepackage{iopams,epsf,graphicx}
\begin{document}

\title{On mobility of electrons in a shallow Fermi sea over a rough seafloor}

\author{Kamran Behnia}

\address{Laboratoire de Physique et d'Etude de Mat\'eriaux (Affiliated to CNRS and UPMC)\\
 ESPCI, 10 Rue Vauquelin, F-75005 Paris, France}

\begin{abstract}
Several doped semiconductors, in contrast to heavily-doped silicon and germanium, host extremely mobile carriers, which give rise to quantum oscillations detectable in relatively low magnetic fields. The small Fermi energy in these dilute metals quantifies the depth of the Fermi sea. When the carrier density exceeds a threshold, accessible thanks to the long Bohr radius of the parent insulator, the local seafloor is carved by distant dopants. In such conditions, with a random distribution of dopants, the probability of finding an island or a trench depends on on the effective Bohr radius, a$_{B}^{*}$ and the carrier density, n. This picture yields an expression  for electron mobility with a random distribution of dopants: $\mu_{RD}\propto $(a$_{B}^{*})^{1/2}$ n$^{-5/6}$, in reasonable agreement with the magnitude and concentration dependence of the low-temperature mobility in three dilute metals whose insulating parents are a wide-gap (SrTiO$_{3}$), a narrow-gap (PbTe) and a ``zero''-gap (TlBiSSe) semiconductor.
\end{abstract}



\section{Introduction}
In an outreach paper written in the heyday of elemental Fermiology, Mackintosh remarked that while ``few  people  would  define a metal as a ‘solid  with a  Fermi  surface’, this may  nevertheless  be  the  most  meaningful definition of a metal.''\cite{mackintosh1963}. In opposition to a metal, a semiconductor can be defined as a solid deprived of a Fermi surface.

Doping a semiconductor\cite{shklovskii1984}, can eventually turn it to to metal. This metal-insulator transition\cite{mott1990} has been a central issue of the condensed-matter physics during the second half of the twentieth century. The most-explored system has been silicon doped with phosphorus, its immediate neighbor to the right in the periodical table. Experiment has identified a sharp transition and has accurately measured the critical concentration of dopants for emergence of metallicity\cite{rosenbaum1980}. Above this threshold concentration, silicon becomes metallic: It conducts electricity at zero-temperature. This is an alternative definition of a metal\cite{edwards2010} and more inclusive than the one mentioned above. Historically, the debate on metal-insulator transition have been focused on ``weak metallicity'', which only requires the presence of mobile electrons at zero temperature and not ``strong metallicity'', which implies the existence of a well-defined Fermi surface. In a variety of insulators, the critical concentration of dopants for metal-insulator transition has been identified by experiment \cite{edwards1978} and the results match the expectations of the Mott criterion for metal-insulator transition\cite{mott1956,mott1961}, which identifies the effective Bohr radius of the host insulator as the key parameter in a given material.

\begin{figure}
\begin{center}
\includegraphics[width=14cm]{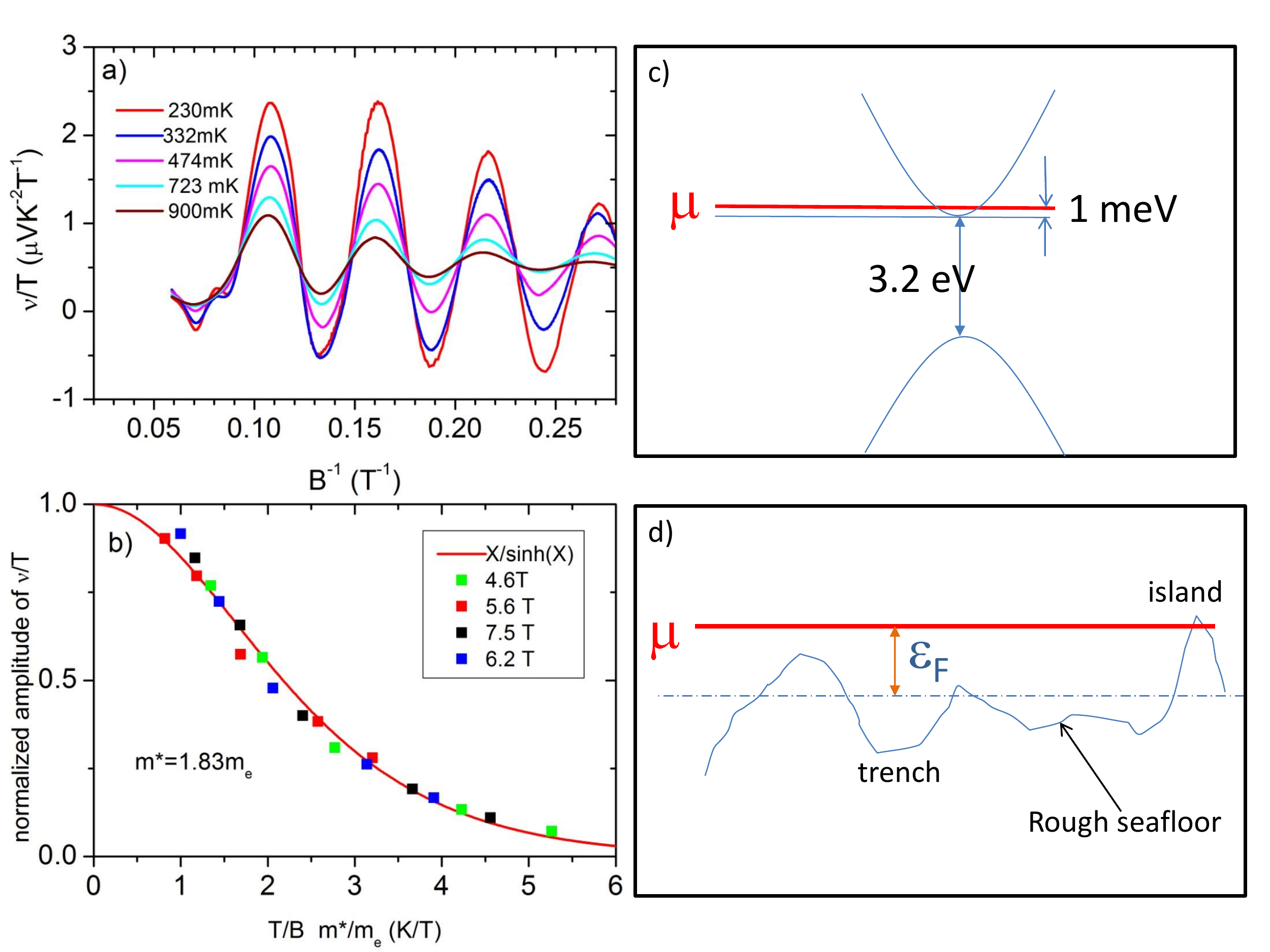}
\caption{Quantum oscillations of the Nernst coefficient (a) and the temperature dependence of the amplitude of the oscillations (b) in SrTiO$_{3-\delta}$ [adapted from ref.\cite{lin2013})]. They attest to the existence of a well-defined Fermi surface and quantify the amplitude of the Fermi energy. c) The results imply that the chemical potential is 1meV above the bottom of the conduction band separated from the valence band by a gap of 3 eV. d) The seafloor is this shallow Fermi sea is rough and the low-temperature mean-free-path of electrons is set by the average distance between islands or trenches, which are potential scattering centers.}
\end{center}
\end{figure}

According to the Pauli exclusion principle, electrons of a solid which share the same discrete quantum numbers (such as spin and valley) cannot have the same wave-vector. They will fill k-states in a Fermi-Dirac distribution with an edge separating the filled and empty states.  The sharpness of this edge is set by the inverse of the mean-free-path, which defines the size of pixel in the k space. A sharp Fermi surface has an experimental signature other than a finite zero-temperature electric conductance.  Landau quantification truncates a Fermi surface to a number of concentric tubes leading to quantum oscillations in various physical properties\cite{shoenberg1984}. In many doped semiconductors,  such oscillations have been reported decades ago. The list includes n-doped strontium titanate\cite{gregory1979,uwe1985,lin2013,allen2013,lin2014}, doped IV-VI semiconductors such as PBS \cite{stiles1961}, PbTe\cite{stiles1961,burke1970,jensen1978} and SnTe\cite{burke1965,savage1972}. This is also the case of heavily-doped germanium\cite{bernard1964,nakamura1967,carrere1974}. A recent study of Nernst effect in SrTiO$_{3}$ found quantum oscillations of a very large magnitude with a single and well-defined frequency (See Fig.1). This result implies that this wide-gap semiconductor, after loosing one out of its 10$^5$ oxygen atoms, becomes a metal in the strongest sense of the word: a solid with a Fermi surface.

From this perspective, one unexpected finding of the recent research on topological insulators is instructive. The quest to find ultra-mobile electrons on the surface protected from backscattering mostly documented \emph{bulk} mobile electrons giving rise to easily-detectable quantum oscillations\cite{butch2010,cao2012,fauque2013,lahoud2013}. It is now clear that topological insulators are often on the metallic-side of the metal-insulator transition as a consequence of uncontrolled doping and a long Bohr radius in the bulk. The irony was not pursued however. Bulk electrons were often treated as a nuisance, impeding topological ``insulators'' to qualify as true insulators. Their proper mobility, with notable exceptions\cite{skinner2012}, was not put under scrutiny.

The subject of the present paper is the intrinsic limit to carrier mobility in a dilute metal made by the introduction of a random distribution of dopants in a semiconductor with sufficiently long Bohr radius. A pattern is visible in the available experimental data. In these systems, at identical carrier density, mobility of electrons is much higher  than in metallic silicon or metallic germanium. Moreover, the variation with carrier concentration is quite distinct. The main argument of this paper is that both the magnitude of the mobility and its density dependence  can be understood by conceiving the simplest model for the probability of finding a scattering center in a shallow Fermi sea. The key parameter remains the Bohr radius of the host insulator, which by setting the Thomas-Fermi screening length, shapes the local seafloor of the Fermi sea.

\section{Carrier mobility in  metallic semiconductors}

\begin{figure}
\begin{center}
\includegraphics[width=14cm]{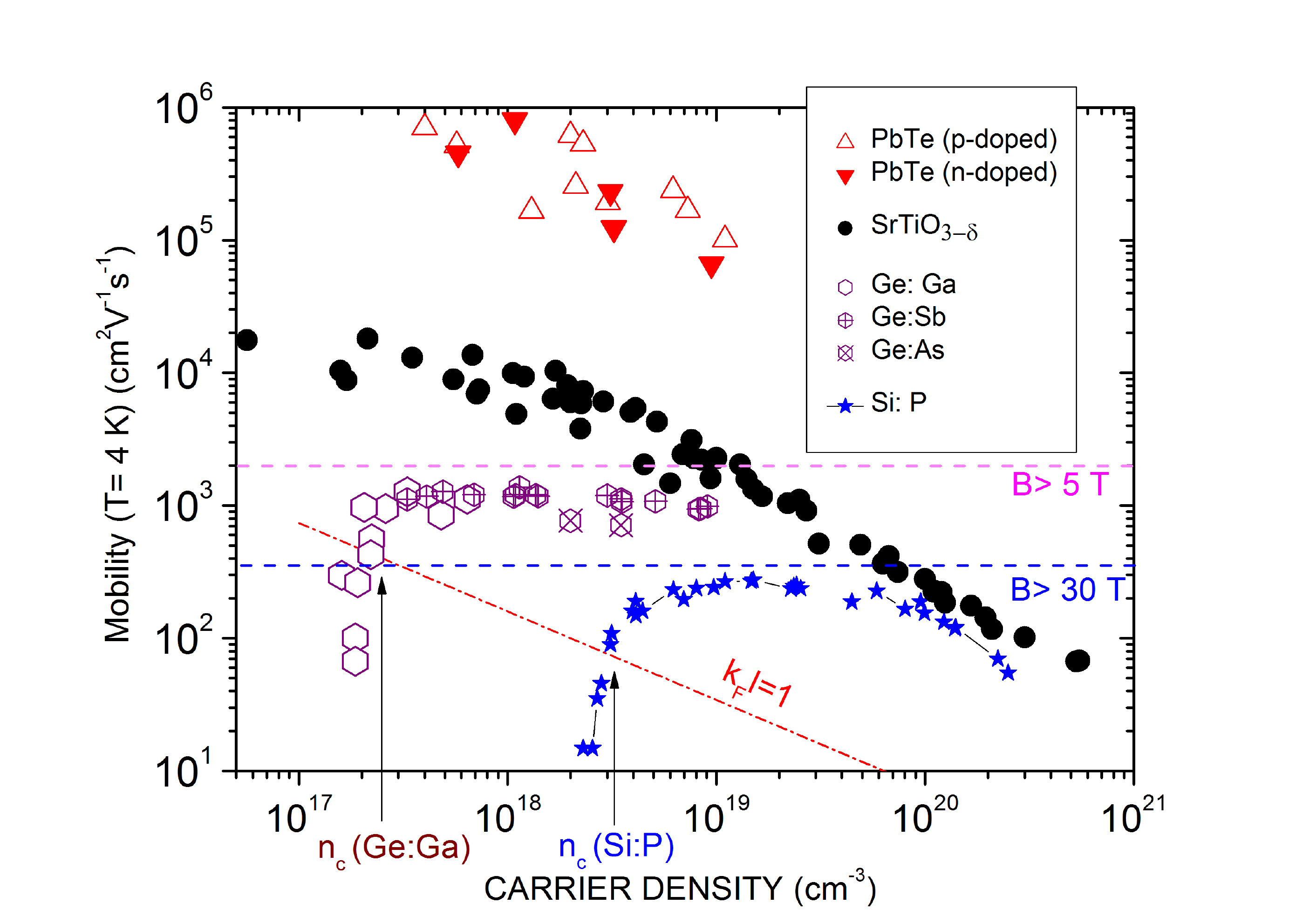}
\caption{Mobility as a function of carrier density in a number of doped semiconductors. Neither doped silicon, nor doped germanium show a strong variation in mobility, when the carrier density is above the threshold of metal-insulator transition. In both cases, the metal-insulator transition and the onset of localisation (corresponding to k$_F \ell \sim$ 1 occurs almost simultaneously. On the other hand, in PbTe and SrTiO$_{3-\delta}$, which have both a long Bohr radius,  mobility enhances with decreasing carrier concentration and localization is easily avoided. Solid horizontal lines mark the required mobility to see detect quantum oscillations at a given magnetic field.}
\end{center}
\end{figure}

Fig. 2 Compares the  Hall mobility at liquid-Helium temperature of of four different semiconductors as a function of carrier concentration between $5\times10^{16}cm^{-3}$ and $10^{21}cm^{-3}$. The reported data for phosphorus-doped silicon\cite{yamanouchi1967}, arsenic-doped\cite{katz1965}, antimony-doped\cite{sasaki1975} and gallium-doped\cite{watanabe1998} germanium are presented together with those for p-doped\cite{jensen1978,allgaier1958} and n-doped\cite{allgaier1958} PbTe, as well as for oxygen-deficient SrTiO$_{3}$ \cite{lin2014,frederikse1967,koonce1967,spinelli2010}. The figure reveals several remarkable features.

\begin{table}
\centering
\begin{tabular}{c c c c c}
  \hline
  \hline
  SYSTEM & $\epsilon/\epsilon_{0}$ & $m^{*}/m_{e}$ & a$^{*}_{B}$ (nm) \\
   \hline
  Si & 12.5 & 0.45 & 1.5 \\
   \hline
  Ge & 16 & 0.24 & 3.5 \\
   \hline
  SrTiO$_{3}$ & 20000 & 1.8 & 600\\
   \hline
  PbTe & 1000 & 0.07  & 800 \\
  \hline
\end{tabular}
\caption{Normalised electric permittivity, $\epsilon$, the average effective, $m^{*}$, and the effective Bohr radius (a$^{*}_{B}= \frac{4\pi \epsilon\hbar^{2}}{m^{*}e^{2}}$), in four different semiconductors.}
\end{table}

The first is the magnitude of the low-temperature mobility. The room-temperature mobility in both SrTiO$_{3}$ or PbTe is disappointingly low, closing the door to numerous applications. But as electrons are cooled down the resistivity decreases by several orders of magnitude. The amplitude of inelastic scattering, a remarkable subject by itself\cite{lin2015} will not be discussed here. The focus of the present paper is the upper bound to mobility set by elastic scattering, which as one can see in the figure, is very different among the four semiconductors. At a carrier density of $10^{18}cm^{-3}$, the low-temperature mobility in PbTe is orders of magnitude larger than in SrTiO$_{3}$, which in its turn exceeds by far the mobility in germanium. At this carrier density, silicon is an insulator and its low-temperature mobility has dropped to zero.

There is a second striking feature in the figure. While metallic silicon and germanium present an almost flat mobility in the concentration window just above their metal-insulator transition, in both PbTe and SrTiO$_{3}$, the mobility  clearly varies with carrier concentration. Remarkably, this allows them to keep a large k$_{F} \ell$ in the range under consideration and to avoid localisation. On the other hand, the flat mobility in silicon and germanium makes them vulnerable to localization as k$_F$ is reduced.  In both cases, the onset of metal-insulator transition (which matches the Mott criterion) occurs close to where the condition for Anderson localisation (k$_{F} \ell\sim 1$) is satisfied. As Mott himself remarked in the preface of the 1990 edition of his book\cite{mott1990}, while metal-insulator transition in Si:P occurs at the critical density corresponding to the Mott criterion, it is probably an Anderson transition driven by disorder.

Finally, the figure instructs us on the amplitude of magnetic field required to observe quantum oscillations. Roughly one needs $\mu B\geq 1$ to detect them. It is not surprising that there is no trace of a report on quantum oscillations in heavily-doped silicon, given its mobility. The large electron mobility in dilute PbTe is the reason that an electromagnet providing a field as small was sufficient for high-resolution fermiology\cite{burke1970,jensen1978}. In the case of STiO$_{3}$, quantum oscillations become more pronounced as carrier concentration is reduced and the insulating state is approached\cite{lin2013}. This would only look like a paradox if one forgets the density dependence of carrier mobility.

What distinguishes SrTiO$_{3}$ and PbTe from silicon and germanium is their exceptionally long Bohr radius. The relevant parameters for each system are summarized in Table 1. A Bohr radius approaching micron explains why both systems remain strong metals at a carrier density of $10 ^{17} cm^{-3}$, when silicon and germanium have already become (or are still) insulators. The purpose of this paper is to consider the consequences of a Bohr radius deep in the metallic state.

\begin{figure}
\begin{center}
\includegraphics[width=14cm]{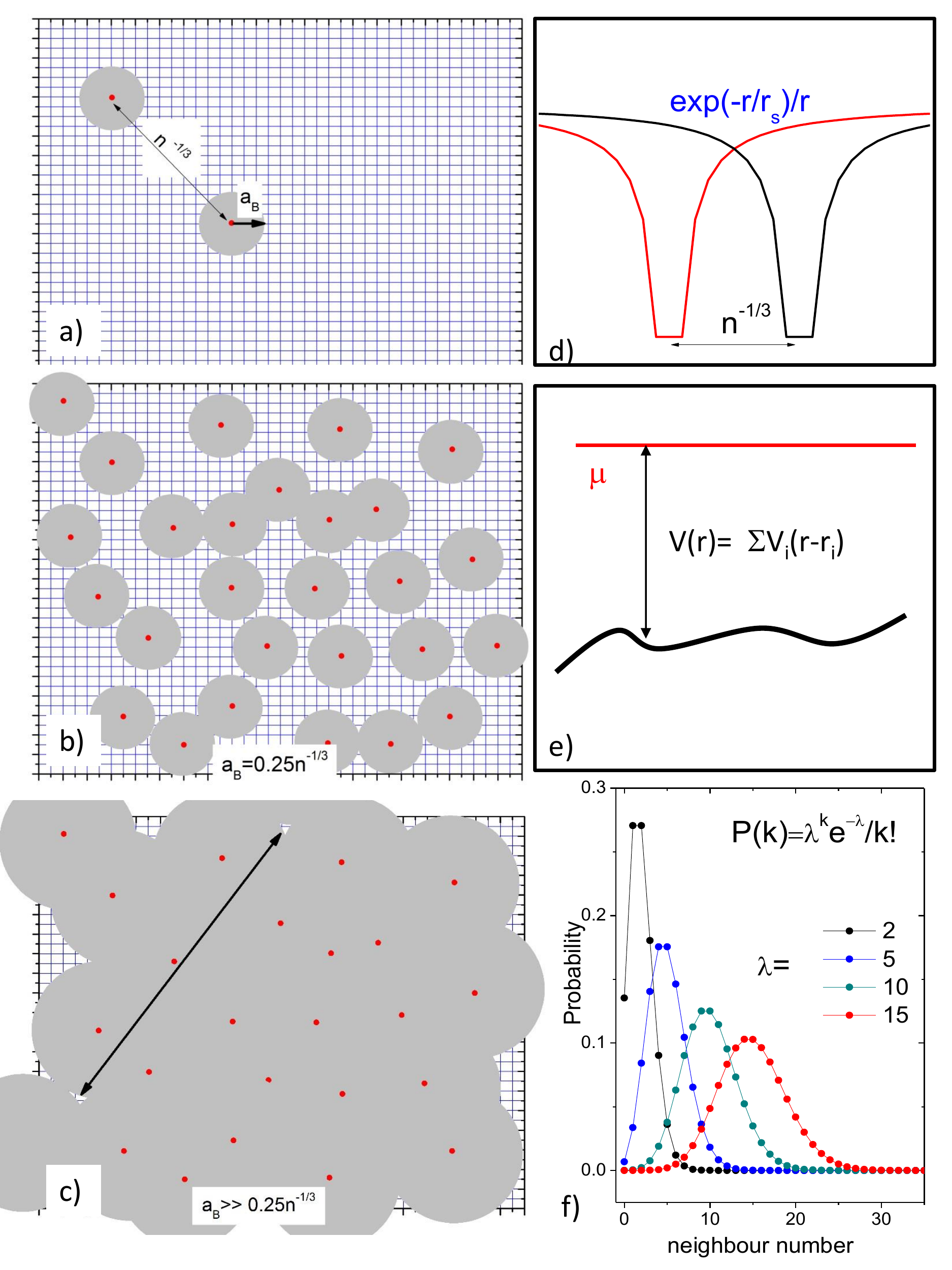}
\caption{a) and b) Illustration of the Mott criterion for the metal-insulator transition. c) When the Bohr radius becomes much longer than the interdopant distance, Bohr spheres interpenetrate. Scattering is governed by statistically unavoidable regions (marked by arrows) in which the local concentration is well below the global average. d) The two relevant length scales in the metallic side of the transition are the screening length, r$_{TF}$, which determine the spatial decay of the potential energy around a dopant and the interdopant distance, $n^{-1/3}$. e) The local energy landscape is set by the sum of all contributions from all potential wells. f) Poisson distribution becomes wider as $\lambda$, the expectancy, increases. For $\lambda >10$, it can be approximated as a normal distribution. }
\end{center}
\end{figure}

\section{Three relevant length scales}
Decades ago, Mott began to draw a picture of how a critical concentration of dopants transforms an insulator to a metal\cite{mott1956,mott1961}. In his argument, three length scales play a key role.

The first one is the effective Bohr radius of the host insulator. The Bohr radius was introduced in the early days of the quantum mechanics as a length scale connecting four different fundamental constants, the Planck constant,$\hbar$, the mass, $m_{e}$,  and the charge, $e$, of electron and the electric permittivity of vacuum, $\epsilon_{0}$:

\begin{equation}
a_B=\frac{4\pi \epsilon_{0}\hbar^{2}}{m_{e}e^{2}}=0.053 nm
\end{equation}

The Bohr radius quantifies the characteristic length scale of the Coulomb interaction in a hydrogen atom. Now, in a crystal, two out of the four parameters can differ from their magnitude in vacuum. The effective mass of quasi-particles depends on their energy dispersion. The electric permittivity (the dielectric constant) is set by the way the medium responds to an applied electric field. The effective Bohr radius reflects these features:

\begin{equation}
a^*_B=\frac{\epsilon}{\epsilon_{0}}\frac{m_{e}}{m^{*}}a_B
\end{equation}

Thus, a large dielectric coefficient and a light effective mass elongate the Bohr radius and enhance the spatial scope of the electric field generated by an ionized dopant.

The second relevant length scale is the Thomas-Fermi screening length, r$_{TF}$, which sets the length scale for the spatial decay of a screened Coulomb (Yukawa) potential, which is expressed as :

\begin{equation}
V(r)=e\frac{exp (-r/r_{TF})}{4\pi \epsilon r}
\end{equation}

Thomas-Fermi screening describes the response of a Fermi-Dirac distribution to a perturbation in charge distribution. Ziman\cite{ziman1960} showed that using the Poisson equation for a perturbative electric potential, one can the following expression for r$_{TF}$ (expressed here in S. I. units):

\begin{equation}
r_{TF}^{2}=\frac{\epsilon_{0}}{e^{2}N(\epsilon_{F})}=\frac{\pi a^{*}_{B}}{4k_{F}}
\end{equation}

Roughly speaking, while a$^*_B$ quantifies the spatial extension of a Coulomb potential well in an insulating medium, r$_{TF}$  is a measure of how spatially fast the well is screened in a metal. The Mott's criterion for metal-insulator transition corresponds to:
\begin{equation}
r_{TF}=a^*_B
\end{equation}

When r$_{TF}$ becomes longer than a$^*_B$, the screening of Coulomb interaction in the metal extends over a distance longer than the size of the well in the insulting medium. Electron-hole pairs become unstable and electrons trapped in adjacent potential wells connect and form a conduction band.  Ashcroft noticed that this is the critical value for the emergence of a bound level in a Thomas-Fermi field and this suffices to make the Fermi sea unstable\cite{ashcroft1993}.

Mott expressed his criterion in terms of the interelectron distance, which is n$^{-1/3}$ in isotropic medium with a density of n. This is our third length scale, which is linked to the Fermi wave-vector in a Fermi-Dirac distribution :

\begin{equation}
n= \frac{k_{F}^{3}}{3\pi^{2}}
\end{equation}

One can use Eq. 2, 4 and 6 to rewrite Eq. 5 and derive the usual expression of the Mott criterion:

\begin{equation}
a^*_B n^{1/3}= \frac{1}{4}(\frac{\pi}{3})^{1/3} \simeq 0.253
\end{equation}

The Mott criterion has been derived in different ways. Using a polarization-potential method, Bhatt\cite{bhatt1981,bhatt1987} showed that in a cubic lattice of hydrogen atoms, the gap between upper and lower Hubbard bands closes when the lattice constant is reduced to approximately four times the Bohr radius. In absence of disorder, the transition is expected to be first order.

Experimental data on a variety of systems indicate that metal-insulator transition in doped semiconductors occur at a critical density close to the Mott criterion\cite{edwards1978,edwards1995}. In the case of silicon and germanium doped with different dopants, the critical density varies as expected with the magnitude of Bohr radius with small, and may be significant, deviations\cite{newman1983}.

Let us recall that when the equality expressed by Eq. 7 is fulfilled, metallicity emerges thanks to percolation. Yet many Bohr spheres do not touch each other (Fig.3b). For higher dopant concentration and/or much longer Bohr radius, we reach a regime with many Bohr spheres interpenetrating each other (Fig. 3C). Let us examine the consequences.

\section{Connection to the rough Fermi seafloor}
Consider a lattice with a lattice parameter of a$_{0}$, hosting a random distribution of wells. At an occupied site $i$, the local potential can be expressed as:
\begin{equation}
r \leq a_{0} \Rightarrow V_{i}=V_{0}\newline
r > a_{0} \Rightarrow V_{i}=V_{0}a_{0}\frac{exp(-r/r_{TF})}{r}
\end{equation}

Here, V$_{0}$ quantifies the maximum depth of the potential well, which can be assimilated to an isolated impurity level. Now, when there is a density $n$ of such wells,  the local potential level is set by the contribution of all occupied sites, near or far. The average potential level would be:

\begin{equation}
<V_{loc}>=V_{0}a_{0}n\int_{0}^{\infty}\frac{exp(-r/r_{TF})}{r} 4\pi r^{2}dr= 4\pi V_{0} a_{0}nr_{TF}^{2}
\end{equation}

\begin{table}
\centering
\begin{tabular}{c c c c}
  \hline
  \hline
  SYSTEM & a$_{0}$ (nm) & a$^{*}_{B}$ (nm)  & n$_{c2}(cm^{-3}$) \\
   \hline
  Si & 0.54 & 1.5 & $ 7.5\times 10^{20}$\\
   \hline
  Ge & 0.57 & 3.7 &  $1.8\times 10^{20}$\\
   \hline
  SrTiO$_{3}$ & 0.39 & 600 & $1.5\times 10^{17} $\\
   \hline
  PbTe & 0.65 & 800 & $ 4.6\times 10^{16}$ \\
  \hline
\end{tabular}
\caption{The second critical doping estimated using, the lattice parameter and the effective Bohr radius of the four semiconductors. Above this threshold density, the local potential level is mostly set by contribution of global potential wells.}
\end{table}

The main new idea is that there can be threshold dopant density beyond which $<V_{loc}>$ exceeds V$_{0}$. When this happens, the global contribution of potential wells weighs more than a single contribution event at the dopant site. In this regime, the global contribution shapes the local potential landscape.  Such a situation occurs when the concentration attains a threshold value. Let us call this threshold n$_{c2}$. It is given by expression:

\begin{equation}
a_{B}^{*}a_{0}n_{c2}^{2/3}=(\frac{3}{\pi}^{2})^{1/3}
\end{equation}

Note that this is expression is specific to three dimensions. A similar expression can be obtained for two dimensions by replacing $4\pi r^{2}dr$ with $2\pi rdr$ in the integral of Eq. 9.

Table 2 gives the magnitude of n$_{c2}$ in the four semiconductors under discussion assuming that a$_0$ is the actual lattice parameter of the semiconductor. As seen in the table,  in both PbTe and SrTiO$_{3}$, the magnitude of n$_{c2}$ is much lower than in silicon and germanium. It is tempting to attribute the striking difference in the density dependence of mobility seen in Fig. 2 to the fact that in contrast to the two elemental semiconductors, the long Bohr radius in PbTe and SrTiO$_{3}$ allows them to easily attain a regime were the local potential landscape is traced by the contribution of many potential wells.

In a metal, where electrons can freely move, the chemical potential is everywhere the same but the local depth of the Fermi sea is not flat, because the environment created by dopants is random and not periodic (Fig. 3e).   This real-space picture has its counterpart in the momentum space. The mobile electrons have a wave-vector and occupy a band originating from orbitals provided by atoms in the stoichiometric host.  Therefore, there is a Fermi surface with a well-defined locus in the Brillouin zone ($\Gamma$-point in the case of SrTiO$_{3}$ and L-point in the case of PbTe). The roughness of the seafloor reflects the stochastic distribution of dopants, which have generated metallicity. The distribution is assumed to be a Poisson distribution\cite{bhatt1987}, an assumption backed by experimental evidence\cite{thomas1981}.

In order to find the simplest model to describe the electron mobility in such a Fermi sea, we begin by recalling the simple expression linking the mobility of electrons to their mean-free-path, $\ell$:

\begin{equation}
\mu=\frac{e}{\hbar}\frac{\ell}{k_{F}}
\end{equation}

Now, what would be the intrinsic and elastic mean-free-path of electrons in such a context? This question can be replaced by another one: what is the average distance between two scattering centers in such a Fermi sea? A scattering center is where the (statistically unavoidable) deviation from the average density of dopants creates an island or a trench in the Fermi sea. Let us consider the mean-free-path divided by the interatomic distance $\frac{\ell}{a_{0}}$ which is a dimensionless number.  How many atomic sites shall one travel to find a place where the global contribution of dopants is such that there is a significant deviation from the average depth of the Fermi sea. In the simplest conceivable model, this number scales with two other dimensionless numbers, the interdopant distance and the rate of variation of a single potential well. Therefore:

\begin{equation}
\frac{\ell}{a_{0}}=C \frac{n^{-1/3}}{a_{0}}\frac{r_{TF}}{a_{0}}
\end{equation}

In other words, this distance is simply proportional to the distance between dopants, n$^{-1/3}$ and to the screening length, r$_{TF}$. The latter amplifies the scope of the well created by each extrinsic atom. C is a dimensionless parameter expected to be of the order of unity. Injecting the relevant expressions for n$^{-1/3}$ and r$_{TF}$, one finds an expression for intrinsic mobility in a Fermi sea generated from a random distribution of dopants.

\begin{equation}
\mu_{RD}=\frac{C}{\sqrt{12\pi} a_{0}} \frac{e}{\hbar }(a_{B}^{*})^{1/2} n^{-5/6}
\end{equation}

According to this expression, mobility decreases with increasing carrier density in a given system. In different systems at the same carrier density, the longer the Bohr radius, the larger the expected mobility. Note that we are focusing on the mobility of isotropic three-dimensional crystals when their carrier density is larger than n$_{c2}$ as defined by Eq. 10.

\begin{figure}
\begin{center}
\includegraphics[width=12cm]{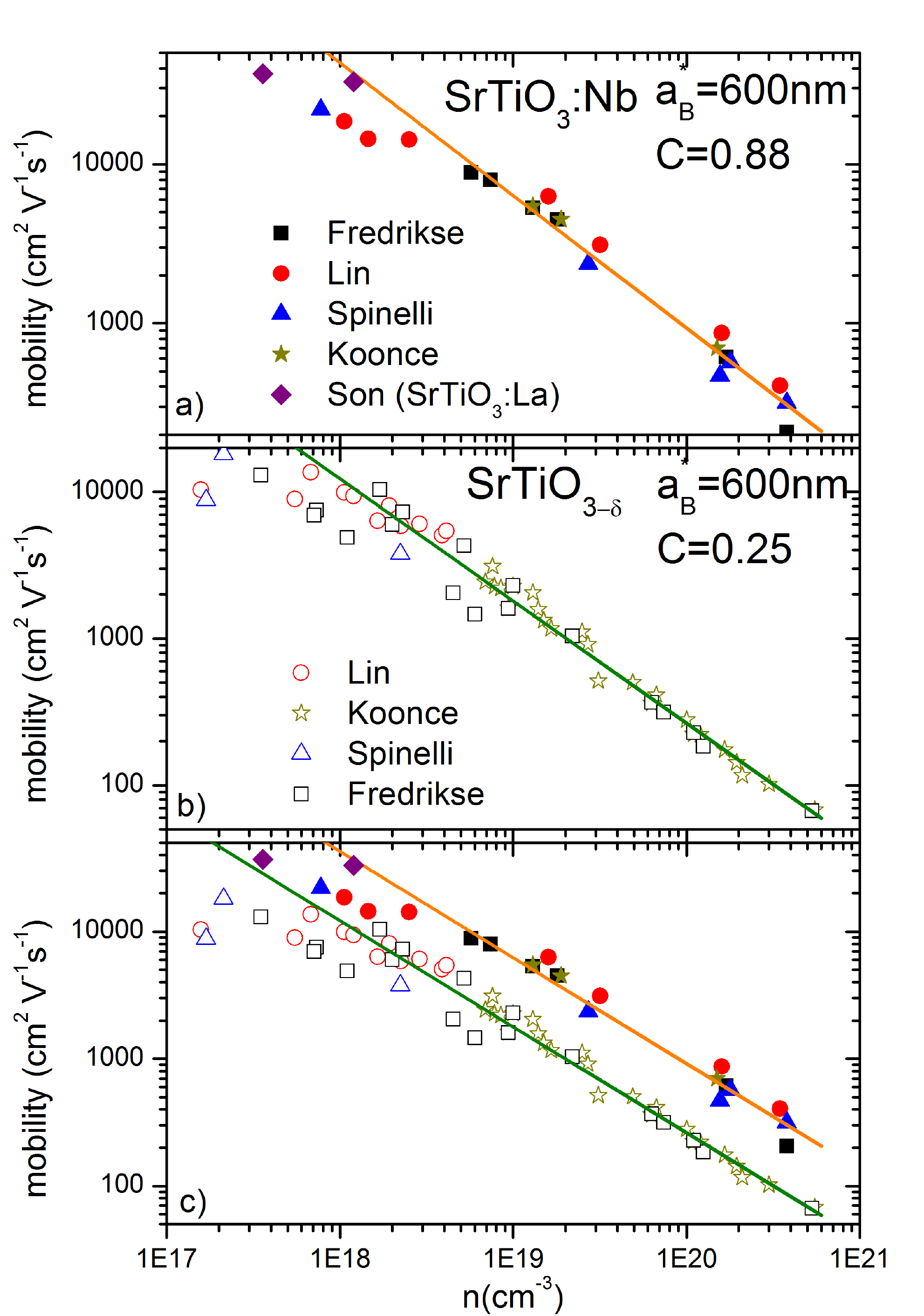}
\caption{Mobility as a function of carrier density in Nb-doped (top) and reduced (middle) SrTiO$_{3}$. The experimental data are reported in references \cite{frederikse1967,koonce1967,spinelli2010,lin2014,jon2010}. Two data points represent the data reported for La-doped SrTiO$_{3}$\cite{jon2010}.  Solid lines represent the expected variation of $\mu_{RD}$ according to Eq. 13.  In the bottom panel they are compared together. }
\end{center}
\end{figure}

\section{Comparison with experimental data}

\begin{figure}
\begin{center}
\includegraphics[width=12cm]{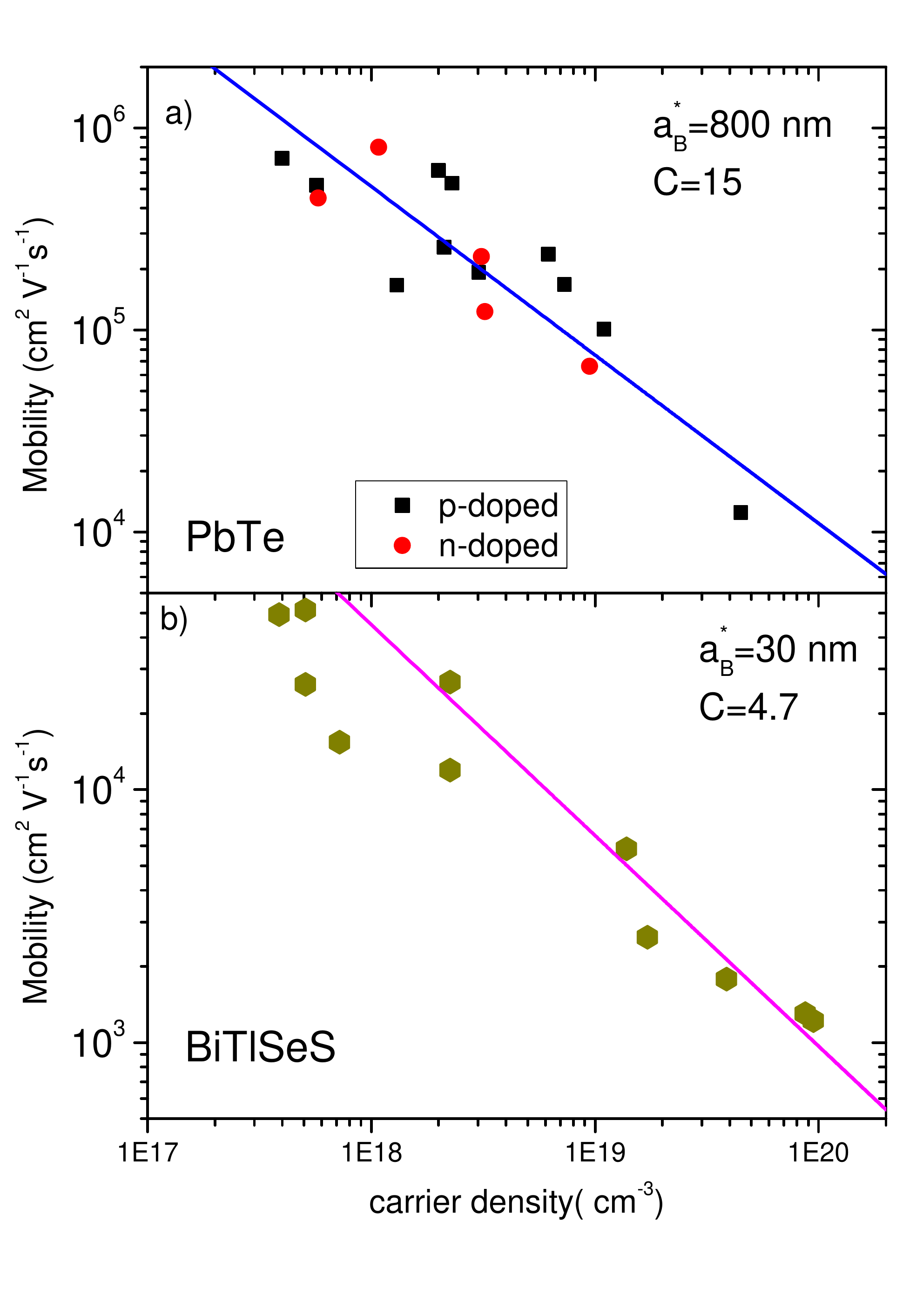}
\caption{Mobility as a function of carrier density in PbTe (top)\cite{allgaier1958,jensen1978}  and TlBiSSe (bottom)\cite{novak2015}. Solid lines show the expected variation of $\mu_{RD}$ according to Eq. 13.}
\end{center}
\end{figure}

 Fig. 5 presents the mobility data reported in oxygen-deficient and Nb-doped SrTiO$_{3}$ by four different groups\cite{frederikse1967,koonce1967,spinelli2010,lin2014}. No disagreement is visible between the data reported by different groups. For comparison, we have also included the  data reported on La-doped SrTiO$_{3}$\cite{son2010}, which presents a mobility slightly larger than Nb-doped samples in the dilute limit. As seen in figure, the expression for $\mu_{RD}$ gives a reasonable account of the data.  The adjustable parameter C is roughly three times lower in the oxygen-deficient case. We will come back to this.

Fig. 6 presents a similar treatment of the available data on PbTe\cite{allgaier1958,jensen1978} and BiTlSeS, which was the subject of a recent study. Novak\emph{ et al.}\cite{novak2015} quantified  mobility as a function of concentration in a system with a vanishing gap tuned half-way between BiTlSe$_{2}$ and BiTlS$_{2}$, two band insulators (one trivial and one topological) with  gaps of comparable amplitudes\cite{xu2011}. The light cyclotron mass(m$^{*}$=0.14m$_{e}$) combined to a relatively large dielectric constant($\epsilon \sim 80\epsilon_{0}$) yields an effective Bohr radius of a$_{B}^{*}\sim~$30 nm.  As seen in Fig. 6, the expression for $\mu_{RD}$ yielded by Eq.13 seems a satisfactory description of the experimental data. Note the amplitude of parameter C, which is large compared to SrTiO$_{3}$.

Note that in this crude treatment, we have neglected the fact that neither the effective mass nor the dielectric constant do not stay constant as the carrier density is modified by several orders of magnitude. The apparent success of the fit indicates that the principal ingredients are present in the above-mentioned picture. Note that a blind linear fit of the logarithm of mobility vs. logarithm of density yields an exponent, which is somewhat lower than -5/6) (See table 3). Note that in all cases a downward deviation from the power-law behavior is visible at the lowest concentration. This is not surprising as the model is expect to brak down with the approach of n$_{c2}$.

What physical properties set the dimensionless fitting parameter C? To answer this question, let us consider two opposing effects of the change in carrier density on scattering rate. We have neglected them up to this point. The first is set by the comparative depth of the Fermi sea and the scattering potential. As the sea becomes deeper, the deviation from average concentration required for an island to emerge becomes harder to satisfy. Moreover, the deeper the potential well, the easier it is to satisfy this criterion. This effect is proportional to V$_{0}$/E$_{F}$. Second, electrons with a short wave-length are easier to scatter than those with longer ones. The scattering cross section scales with 1/k$_{F}^{2}$. A fortunate accident for the simple model discussed here is that the density dependence of these two tendencies cancel each other out. More quantitatively, they lead to the following expression for C:

\begin{equation}
C= \frac{1}{a_{0}^{2}k_{F}^{2}}\frac{E_{F}}{V_{0}}=\frac{\hbar^{2}}{2m^{*}a_{0}^{2}V_{0}}
\end{equation}

Thus, light  mass and shallow potential  enhance C and, consequently, the absolute value of mobility. Knowing the effective mass and the lattice parameter, Eq. 14 permits one to extract V$_{0}$ from the magnitude of the fitting parameter, C. The results are given in Table 3. As seen in the table, as one would expect,
the depth of potential well is much lower in the small-gap semiconductors. The difference between oxygen-deficient and Nb-doped SrTiO$_{3}$ is also understandable. Substituting Ti with Nb introduces a single electron at the center of an octahedron. Oxygen vacancy, on the other hand, introduces two electrons and distorts two adjacent octahedra, digging a deeper potential well.

\begin{table}
\centering
\begin{tabular}{c c c c c c c}
  \hline
  \hline
  SYSTEM & $\alpha$ & C & a$_{0}$ (nm)  & m*(m$_{e}$) & V$_{0}$ (eV) & E$_{g}$ (eV) \\
   \hline
 SrTi$_{1-x}$Nb$_{x}$O$_{3}$& -0.73 & 0.87 & 0.39 &  1.8& 0.16 & 3.2 \cite{capizzi1970} \\
   \hline
  SrTiO$_{3-\delta}$ &-0.72& 0.25 & 0.39 & 1.8 & 0.56& 3.2 \cite{capizzi1970}\\
   \hline
  PbTe (self-doped)&-0.73 & 15 & 0.65 & 0.07& 0.1 & 0.2 \cite{dimmock1966}\\
  \hline
 BiTlSeS(self-doped)&-0.65 & 4.7 & 0.45 & 0.14 & 0.06 & $<$0.04 \cite{xu2011} \\
   \hline
\end{tabular}
\caption{Parameters of the four systems under scrutiny. $\alpha$ represents the exponent found by fitting the data to $\mu \propto n^{\alpha}$. C is the parameter extracted by fitting the data to Eq. 13. V$_{0}$ is extracted from C, the effective mass, m$^{*}$, the lattice parameter a$_0$ and Eq. 14. E$_{g}$ is the the band gap.}
\end{table}

\section{Summary}
This paper puts under scrutiny an experimental fact already reported but rarely analyzed. In metallic semiconductors with a long Bohr radius, mobility is large and continuously decreases with increasing carrier density. Both these features are absent in silicon or germanium. According to the argument proposed here, this happens because in these metallic semiconductors the global distribution of dopants shapes the local seafloor. For this to happen,  a threshold concentration of dopants is required. This second critical concentration becomes attainable only when the Bohr radius is long enough. With a set of most unsophisticated assumptions, one can formulate an expression for the intrinsic mobility of a metallic semiconductor with random distribution of dopants. In spite of neglecting subtle details, the expression is in good agreement with the data on four different semiconductors.

This work is supported by \emph{Agence Nationale de la Recherche} through the SUPERFIELD project.

\end{document}